\documentstyle[11pt]{article}

\def\be{\begin{equation}}
\def\ee{\end{equation}}

\begin{document}

\title{{\hfill\small\tt Phys.\,Rev.\,{\bf B 45},\,12084-12087\,(1992)}\\
{\hfill}\\
Charge distribution in two-dimensional electrostatics}

\author{I.Kogan\footnote{Permanent address: 
Institute for Theoretical and Experimental Physics, 117259 Moscow, Russia}\\
{\small\em Department of Physics,
University of British Columbia,}\\{\small\em Vancouver, British Columbia,
Canada V6T 1Z1}\\
A.M. Perelomov\footnote{Permanent address: 
Institute for Theoretical and
Experimental Physics, 117259 Moscow, Russia}\\
{\small\em Centre de
Recherches Mathematiques, Universite de Montr\'eal,}\\{\small\em Montr\'eal,
Queb\'ec, Canada H3C 3J7}\\ 
G.W.Semenoff\\
{\small\em Department of Physics,
University of British Columbia,}\\{\small\em Vancouver, British Columbia,
Canada V6T 1Z1}\\
{\small Received 13 November 1991}
}
 
\date{}
\maketitle

\begin{abstract}\noindent
We examine the stability of
ring-like configurations of $N$ charges on a plane interacting
through the potential $V(z_1,\dots,z_N)= \sum\vert
z_i\vert^2-\sum_{i<j}\ln\vert z_i-z_j\vert^2$. We interpret the
equilibrium distributions in terms of a shell model and compare
predictions of the model with the results of numerical simulations
for systems with up to 100 particles.
\end{abstract}

\noindent
Determining the distributions of charged particles in a central
potential is a classic problem which has been studied since the nature
of the electric force was first understood [1]. In two dimensions the
Coulomb potential varies logarithmically with distance and
describes, for example, the interaction of parallel charged wires.
One could consider the physical problem of determining the
electrostatic configuration of a group of parallel
wires with like charges confined by a central force.

Ideas such as this have been applied to the description of the
`crystalline state' of cooled particle beams [2], where, in their rest
frame, the beams are approximated by line-charges, and the central
force, implemented by focusing magnets for example, is imposed to
maintain the width of the beam.  A similar situation arises in systems
of charged particles where the dielectric properties of their
environment are so highly anisotropic that the system is approximately
two dimensional.  An example is the the distribution of charged ions
on superfluid surfaces.  The Coulomb interactions in this system can
be made effectively logarithmic and the crystalline states of the ions
have been studied [3].

Another setting where Wigner crystal states are thought to play a role
are the low density and high field states in the fractional quantum
Hall system [4].  In the incompressible quantum fluid which exhibits
the fractional quantum Hall effect, the ground state is described
by Laughlin's wavefunction [5,6].  At higher fields and lower
densities of electrons, it is conjectured that the same system is in a
Wigner crystal state. (This state has recently been observed
experimentally [4].)  There is the interesting question of
whether there can be other states intermediate between a quantum Hall
state and a Wigner crystal and whether these states could be described
by a Laughlin-like wavefunction.  It is not known to what extent Laughlin's
wavefunction continues to give a reasonable description of the
electronic ground state for low densities and high fields.

We shall begin by studying the electrostatic problem of finding
the configuration of particles which minimizes the
potential energy
\begin{equation}
 V(z_1,\ldots,z_N)=\sum_{i=1}^N\vert z_i\vert ^2-\sum_{i<j}
\ln | z_i-z_j|^2, \end{equation}
where $z_i$ are the complex coordinates of particle positions in a
plane. The central well is the interior potential of a uniformly
charged disc.  It obeys the Poisson equation $-\nabla^2|
z|^2=-4$ so the background charge density is $-1/\pi$, independent
of the radius of the disc.  There is also a repulsive Coulombic (in two 
dimensions)  interparticle potential.  The potential (1) is the (suitably
rescaled) logarithm of Laughlin's wavefunction for the fractional
quantum Hall effect states [4,5].  The wavefunction is given by 
\[
 \psi(z_1,\ldots,z_N)=\prod_{i<j}(z_i-z_j)^\alpha \,\exp\left( -H\sum_i 
\vert z_i\vert ^2/2\right) \]
and $\rho(z_1,\dots,z_N)=\psi^{\dagger}\psi=
\exp [-\alpha V\,(\sqrt{H/2\alpha}\,z_i)]$, where 
$H$ is the magnetic field and $\nu={1/\alpha}$ is the filling factor.  
For values of $\alpha$ and $H$ relevant to the fractional quantum Hall
effect the particle distribution and density correlations described by
$\psi(z_1,\dots,z_N)$ are those of an incompressible liquid [6].  As $\alpha$
and $H$ are increased there is a phase transition to a state where  the
probable distributions of particles described by $\psi(z_1,\dots,z_N)$ is
concentrated at the classical minima of $V$.  This is the analog of a
crystalline state.

Let us begin with a simple mathematical problem.  If we have a few
particles in this system, we expect that they will lie on a ring at
some equilibrium radius $R$.  If we add more particles we expect that
they increase the size of the ring to a maximum.  An interesting
question is: how large can the ring be before it is unstable?  Also,
even if the ring is stable to small oscillations, there is the more
difficult question of whether it is actually a global minimum of the
potential energy or whether there exist other, more favorable states.

The symmetry of the problem indicates that the configuration with
$N$ particles lying on the ring is a stationary point of the
potential energy. To reason that a ring should have some maximum
size where it is no longer a local minimum of the energy, consider
the following continuum argument: Instead of point particles we
allow the charge density to be continuous so that the energy is
now given by
\begin{eqnarray*}
V[\rho] & =&\int d^2z\,\vert z\vert^2\,\rho(z) \\
&-&{1\over2}\,\int d^2z\int d^2z'\,\rho(z)\,\rho(z')\ln\vert z-z'\vert^2 \\
&+&\lambda\left( N-\int d^2z\,\rho(z)\right).\end{eqnarray*}
Here $\lambda$ is a Lagrange multiplier to fix the total charge. 
This potential is stationary where
\[
 \vert z\vert^2-\int d^2z'\,\rho(z')\,\ln\vert z-z'\vert^2-\lambda=0,
\qquad \int d^2z\,\rho(z)=N. \]
These equations are solved by
\begin{equation}
\rho(z)=\left\{ \begin{array}{ll} 1/\pi , & \vert z\vert<R \\
0,& \vert z\vert>R\quad \mbox{with}\,\,R^2=N,\end{array} \right. 
\end{equation}
 a uniform disc-shaped charge density (which, as expected, exactly 
compensates the background charge of the disc). Thus we expect that, 
in the limit of large numbers of particles, where we can approximate 
their distribution as continuous, the average of the distribution is 
uniform rather than ring-like and therefore large rings should be 
unstable unless they contain a sufficiently large charge in their interior.

In the following, we shall consider (and find an exact answer for) the
slightly more general question of how large the ring can be when there
is an azimuthally symmetric charge distribution inside.  Such a
configuration should minimize the energy 
\begin{eqnarray}
V_Q(z_1,\dots,z_N) & =&\sum_i \vert z_i\vert^2-\sum_{i<j}\ln\vert
z_i-z_j\vert^2 \nonumber \\
&+& \sum_i\int d^2z'\,\ln\vert z_i-z'\vert^2\,\rho(\vert z'\vert) ,
\end{eqnarray}
where $\int_{\vert z'\vert<R} d^2z' \,\rho(\vert z'\vert)=Q$.

The first variation of $V_Q$ is
\begin{eqnarray}
&&\delta V_Q(z_1,\ldots,z_N) \nonumber \\
&=&\sum_i\left( {\bar z}_i\,\delta z_i+\delta{\bar z}_i\,z_i\right) 
-\sum_{i<j}\left( {{\delta z_i-\delta z_j}\over {z_i-z_j}}+
{{\delta{\bar z}_i-\delta{\bar z}_j}\over{{\bar z}_i-{\bar z}_j}}\right) 
\nonumber \\
&-&\sum_i\int dz' \left( {{\delta z_i}\over {z_i-z'}}+{{\delta{\bar z}_i}
\over {z_i-z'}}\right) \rho(\vert z'\vert) .\end{eqnarray}
We make the ansatz $z_k=R\exp(2\pi ik/N)$ for $k=1,\ldots,N$. 
This assumes that $N$ particles sit on a ring which encloses an 
azimuthally symmetric charge distribution.  Of course, if there are 
particles inside the ring, the charge distribution inside will not be
azimuthally symmetric.  However, especially in the limit of large
numbers of particles, cancelling forces form approximately evenly
distributed point charges should make this a good approximation.

From (4) we obtain the equation 
\begin{equation} 
R^2=Q+\sum_{k=i}^N {1\over {1-\exp (2\pi ik/N)}}=Q+(N-1)/2, \end{equation}
where we have used the sum rule (A.2).
Thus, we find that the ring configuration is always an {\em extremum}
of the energy.  In order to see whether it is a {\em local minimum} or
a {\em saddle point} we must compute the eigenvalues of the stability
matrix given by the second variation
\begin{eqnarray}
&&\delta^2 V_Q(z_1,\ldots,z_N)\nonumber \\
& =&2\sum_k \delta {\bar z}_k\,z_k+{1\over2}\sum_{j<k}
\left( {{(\delta z_j-\delta z_k)^2}\over {(z_j-z_k)^2}} +
{{(\delta {\bar z}_j-\delta {\bar z}_k)^2}\over {({\bar z}_j-{\bar z}_k)^2}}
\right) \nonumber \\
&+&\sum_k \int d^2z'\,\rho(\vert z'\vert)\left( {{\delta z_k\,\delta z_k}
\over {(z_k-z')^2}}+{{\delta{\bar z}_k\,\delta{\bar z}_k}\over {({\bar z}_k-
{\bar z}')^2}}\right) .\end{eqnarray}
Using the summation formula in equations (A.2) and (A.3), the
parameterization of the equilibrium positions $\delta z_k=
z_k\left( \delta\,\ln R_k+i\delta\phi_k\right) $, and the formula (5) for
$R^2$, we get 
\begin{eqnarray}
\delta^2V_Q &=&\sum_i\left( 4Q+{1\over6}\,(N-1)(11-N)\right) \delta\ln R_i\,
\delta\ln R_i \nonumber \\
&+&{1\over2}\sum_{i\neq j} {{\delta\ln R_i\,\delta\ln R_j}\over 
{\sin^2\pi(i-j)/N}} +\sum_i{1\over 6}\,(N-1)^2\,\delta\phi_i\,
\delta \phi_i\nonumber \\
&-&{1\over2}\sum_{i\neq j} {{\delta\phi_i\,\delta\phi_j}\over {\sin^2\pi\,
(i-j)/N}}\,.\end{eqnarray}
The stability matrices are related to spectra of the Calogero model 
and can be diagonalized by techniques developed in Ref. 7 and reviewed
in Appendix.  For the angular fluctuations, $\delta\phi_i$, the
spectrum is $ s(N-s)$,  $s=0,\ldots,N-1$, and for radial fluctuations, 
$\delta\ln R_i$ it is $ 4Q+2(N-1)-s(N-s)$, $s=0,\ldots,N-1$. 
The angular modes are non-negative, indicating 
stability to angular fluctuations for all $N$.  The zero mode for 
$s=0$ is a consequence of rotation invariance of $V_Q$.  For large 
enough $N$ some radial modes are negative, indicating instability of 
the radial fluctuations.  The minimum of the radial spectrum occurs at 
$s=N/2$ (if $N$ is even).  The maximum value of $N$ for which this 
minimum is positive is the largest integer less than 
\begin{equation} 
N_{\rm max}=4\left( \sqrt{Q+1/2}+1\right) .\end{equation}
For a system with a total of $M=N+Q$ particles, the largest number which will 
lie on the outside ring is given by the largest integer less than 
\begin{equation}
N_{\rm max}=4\left(\sqrt{M+1/2}-1\right) .\end{equation}
It is interesting to note that, as particles are added to a ring,
the first mode of instability to radial fluctuations appears for the
mode of maximum frequency, i.e. that where every second particle moves
inward and the other particles move outward.  Locally, instead of the
ring rejecting the last particle and forcing it to the center of the
distribution, its first tendency is to split into two rings of roughly
equal size.

A numerical calculation using simulated annealing Monte Carlo methods 
can be used to find the equilibrium distribution of the particles for
$N$ up to 100.  The result is that, to a good approximation,
particles lie on concentric rings with the number of particles per
ring increasing like the square root of the radius of the ring and
with average spatial density close to the value $1/\pi$ given in (2).

The structure that we see is reminiscent of the shell model of the
atom. We can devise a model for predicting the number of particles in
each ring.  We begin by using Eq.(9) to calculate the maximum
number of particles which fit in the outer ring, which depends on the
total number of particles.  Then we subtract that number from the
total and compute how many particles will fit in the next ring given
the total remaining number of particles and so on until the total
number of particles is exhausted.  This gives the maximum occupation
numbers of concentric rings.

There are two limits to the accuracy of this model.  First, the
internal charge distribution is approximated as azimuthally symmetric,
rather than distributed at points.  In the real system the rings are
perturbed by the inhomogeneities of the charge distribution
and are not exactly circular.  We expect that azimuthal symmetry is a
good approximation when the number of particles is large.  Second, 
the model is accurate only when each ring tends to fill to its locally
stable configuration with maximum number of particles.  In almost
every case, this is unlikely as there can be many preferred, lower
energy states where rings are not filled to maximum capacity.

We know that this already happens for a six particles.  Our
theoretical computation indicates that six particles sitting on the
corners of a hexagon is stable to small perturbations.  However,
explicit calculation reveals that the configuration with five
particles sitting on corners of a pentagon with a single particle at
the center is also stable to perturbations and has slightly lower
energy than the hexagon.  Therefore, already for six particles our
shell model is approximate.

It is then interesting to ask how accurate it is for higher numbers of
particles. Some results of a numerical simulations compared with
predictions of the shell model are 
\[ 
\matrix{M: &2&3&4&5&6&7&8&9&10&15&25&100\cr
N_{\rm exp}: &2&3&4&5&5/1&6/1&7/1&7/2&8/2&11/4&14/9/2&31/25/19/14/8/3\cr
N_{\rm th}: &2&3&4&5&6&6/1&7/1&8/1&8/2&11/4&16/8/1&36/28/20/12/4\cr } \]
where, for both the experimental and theoretical values we also denote
the number of particles in the inner rings, starting from
the largest.  We see qualitative agreement of the results of the
model.  However quantitative predictions are reliable only within
about 30\%.  Also, since the rings do not fill to their maximum, 
we tend to underestimate the number of rings\footnote{\,\,The computer 
simulation of the 100 particle case is shown in Fig.1, see the original 
paper}.  We also estimate the accuracy of the computer 
simulation itself to be within about two or three for the population 
of the rings.  (This estimate is obtained from reproducability of the 
results.)

The shells are generally not filled to their maximum population since
global minima of the potential appear first.  It would be interesting
to obtain a ring-filling criterion which sought global minima of the
energy.  This would be an analog of Hund's rule for filling of
electronic orbitals in atoms [8].  We haven't yet succeeded in doing this,
our only present recourse is to explicit calculations and comparisons
of the total energy of different configurations of a few particles and
numerical simulations.

As a test of the accuracy of the shell model, we have used a numerical
simulation to find the equilibrium configurations in the region
between 40 and 60 particles.  Below we show the number and population
of the rings observed and compare with the numbers and populations of
rings which are computed using the shell model, 
$$
\matrix{M:&40&41&42&43&44&45\cr
N_{\rm th}:&21/13/6&21/14/6&22/14/6&22/14/6/1&22/14/7/1&22/15/7/1\cr
N_{\rm exp}:&19/14/6/1&19/13/7/2&18/15/7/2&20/13/8/2&21/15/7/1&20/14/8/3\cr}
$$ $$
\matrix{M:&46&47&48&49&50\cr
N_{\rm th}:&23/15/7/1&23/15/8/1&23/16/8/1&24/16/8/1&24/16/8/2\cr
N_{\rm exp}:&21/14/9/2&21/13/10/3&21/14/9/4&20/14/9/5/1&22/15/9/4  \cr}
$$ $$
\matrix{M:&51&52&53&54&55\cr
N_{\rm th}:&24/16/9/2&24/17/9/2&25/17/9/2&25/17/10/2&25/18/10/2\cr
N_{\rm exp}:&21/16/8/5/1&22/15/9/5/1&22/15/10/5/1&22/16/11/5&21/17/10/6/1\cr}
$$
$$
\matrix{M:&56&57&58&59&60\cr
N_{\rm th}:&26/18/10/2&26/18/10/3&26/18/11/3&26/19/11/3&27/19/11/3\cr
N_{\rm exp}:&23/17/9/6/1&22/17/11/6/1&22/17/11/6/2&23/15/12/5/4&24/16/12/7/1
\cr} $$ 
Experience shows that these numerical calculations are good to
within plus or minus two or three particles per ring. The main source
of error is distortion of the ring by inhomogeneities of the charge
distribution.  This is particularly acute for rings with near maximum
numbers of particles.  We see that our model predicts the population
of the outer shell within 20\% and gives reasonable populations for the 
inner shells.

Thus we see that the shell model describes well the qualitative, and
approximately the quantitative properties of the frozen state of $N$
particles up to 100 or so.

In conclusion, we observe that, even in the case of 100 particles\footnote{\,
\,shown in Fig.1, see the original paper}, 
there is no observable tendency for the systems we 
study by numerical simulations to form a triangular Wigner crystal.
The latter is the expected ground state in the limit of large numbers
of particles.  For a system which is too small, the tendency to form a
regular crystalline state is frustrated by the boundary geometry.  For
large systems, this should be offset by two effects: First, the
boundary energy grows more slowly than the bulk energy (the ratio is
$1/\sqrt{M}$) so eventually the boundary should adjust itself to help
minimize the energy of the bulk.  Second, the effects of the boundary
on particle positions well inside the bulk should be negligible
because of the screening of the long-ranged Coulomb force by the gas
of charged particles themselves.  The screening length in a classical
Coulomb gas is generally a few average spacings.

It is a mystery to us that we do not see the onset of Wigner
crystallization.  We might speculate about possible glass-like or
quasi-crystalline phases of the system and perhaps a phase
transition to other states as we increase the particle number.
(Phase transitions are possible in this system with a finite
number of degrees of freedom because it is classical - there is no
tunneling which could restore a broken symmetry.)  One idea which
supports this speculation is the fact that, for six particles, the
configuration with a pentagon and one particle in the center is
preferred to a hexagon.  Thus, locally the system may have a
tendency to form structures with five-fold symmetry.  However,
this structure cannot form a lattice and so perhaps forms a
frustrated state with no long-range order.  This possibility is
the subject of ongoing investigation. 
\vskip 0.25truein 

\noindent{\bf Acknowledgements.}\,
The authors thank Eric Hiob for helpful conversations and assistance
with the computer simulations. We also thank S.Girvin and P.Stamp
for helpful conversations. This work is supported in part by the Natural 
Sciences and Engineering Research Council of Canada.
\bigskip

\setcounter{equation}{0}
\renewcommand{\theequation}{A.\arabic{equation}} 

\noindent{\bf Appendix.}\, 
The summations and diagonalizations of matrices used in this paper
have been presented previously in the context of the Calogero model [7].  
We shall summarize the technique in this Appendix.
Consider the matrix $L_{jk}=(1-\delta_{jk})\,\left\{ 1+i\cot\pi (j-k)/N 
\right\} $. We can show explicitly that the vector 
$\psi_k^s= \exp (-2\pi iks/N)$, $s,k=1,\ldots,N$, is an eigenvector of $L$, 
i.e.,
\begin{eqnarray}
&& \sum_{k=0,k\neq j}^{N-1}
\left( 1+i\cot\pi(j-k)/N\right) \exp\left(-\,2\pi iks/N\right) \nonumber \\
&=&\exp\left( -2\pi ijs/N\right) \left( \sum_{k=1}^{N-1} {{-2\exp ((2s+1)\pi 
ik/N)}\over {\exp (i\pi k/N)-\exp (-i\pi k/N)}} \right) ,
\end{eqnarray}
where we have combined the terms in (A.1) and changed variables 
$k\to j-k$ in the sum.  After some algebra, the right-hand side of (A.1) is
\[
-\left( \sum_{l=0}^{2s-2}\sum_{k=1}^{N-1}\exp\left[ (2\pi ik/N)\,
(l-s+1)\right] \right) .\]
The eigenvalue can further be written as
\[
\sum_{l=0}^{2s-2}\left( 1-\sum_{k=1}^N \exp [(2\pi ik/N)(l-s+1)]\right) 
=2s-N-1. \]
Thus, the eigenvalues of $L$ are $2s-N-1$, $s=0,\ldots ,N$.

Next, we use the trigonometric identity 
\[
\cot\alpha\,\cot\beta=1-(\cot\alpha-\cot\beta)\,\cot(\alpha-\beta) \]
to show that 
\[ B={1\over2}\left(L^2+2L-{1\over3}\,(n^2-1)I\right) ,\]
where 
\[ B_{jk}=(1-\delta_{jk})\,\left[ 1/\sin^2\pi (j-k)/N\right] .\]
This gives the eigenvalues of the matrix $B$ as 
${1\over2}\,(N^2-1)-2s\,(N-s)$, $s=1,\ldots ,N$ which are used in Eq.(7) 
to get the spectra of small oscillations.  The eigenvalue equation (A.1) 
also gives the sum rule 
\begin{equation} 
\sum_{k=1}^{N-1}{{\exp(-2\pi iks/N)}\over {1-\exp (2\pi ik/N)}}=
s-{1\over2}\,(N+1)\,.\end{equation}
Another sum rule can be derived from the eigenvalue equation for matrix $B$, 
\begin{equation}
\sum_{k=1}^{N-1}{{\cos 2ks\pi/N}\over {\sin^2k\pi/N}}={1\over3}\,
(N^2-1)-2s(N-s)\,.\end{equation}


\begin{thebibliography}{abc}

\vskip 0.3truein
\bibitem[1]{1.} For recent investigation see, for example, T. Tarnai and 
Zs. Gaspar, Proc. R. Soc. Lond. A433 , 257 (1991); M. Calkin, D. Kiang and 
D. Tindall, Am. J. Phys. 55, 157 (1987); 
H. Munera, Nature 320, 597 (1986); B. Van de Waal, Am. J.  Phys. 56, 
583 (1988); M.G. Calkin, D. Kiang and D.A. Tindall, Nature 319, 
454 (1986); A. A. Berezin, Nature 315, 104 (1985); 
O.D. Kellog, {\it Foundations of Potential Theory}, 223, Springer, Berlin, 
1929; D.  MacGowen, Nature 315, 635 (1985); T. W. Melnyk, O. Knop and W. R. 
Smith, Can. J. Chem 55, 1745 (1977) and references therein. 

\bibitem[2]{2.}A. Rahman and J. P. Shiffer, `A condensed state in a system
of stored and cooled ions', {\it Proceedings of Symposium on the
physics of low energy, stored, and trapped particles}, Stockholm,
1987, Phys. Scr. 22, 133-139 (1988); Z. Phys., A331(1), 71 (1988);
Phys. Rev. Lett. 57, 1133 (1986).

\bibitem[3]{3.}F. Gallet, G. Deville, A. Valdes and F. I. B. Williams, Phys. 
Rev. Lett. 49, 212 (1982);

\bibitem[4]{4.}E. Andrei, G. Deville, D. Glatti, F. Williams, E. Paris and B.
Etienne, Phys. Rev. Lett. 60, 2765 (1988).

\bibitem[5]{5.} R.Laughlin, Phys. Rev. Lett. 50, 1395, (1983),

\bibitem[6]{6.} The Quantum Hall Effect, Editors R.E.Prange and S.M.Girvin,
Springer-Verlag, 1987.

\bibitem[7]{7.}F. Calogero and A. M. Perelomov,
Linear Algebra and its Applications 25, 91 (1979).

\bibitem[8]{8.}For a description of Hund rules, see C. Kittel, {\it 
Introduction to Solid State Physics}, J. Wiley and Sons, New York, 
1971, pg. 509.
\end{thebibliography}
\end{document}